\documentclass[12pt,onecolumn,draftcls]{IEEEtran}

\ifCLASSINFOpdf
\else
\fi
\usepackage{graphicx}
\usepackage{epstopdf}
\usepackage{stfloats}
\usepackage{bm}
\usepackage[cmex10]{amsmath}
\usepackage{algorithmic}
\usepackage{algorithm}
\usepackage{array}
\usepackage{mdwmath}
\usepackage{mdwtab}
\usepackage{multirow}
\usepackage{amsfonts}
\usepackage{amssymb}
\usepackage{color}
\usepackage{balance}
\usepackage{makecell}

\begin{document}

\title{Towards Structural Sparse Precoding: Dynamic  Time, Frequency, Space, and Power  Multistage Resource Programming}
\author{Zhongxiang Wei,~\IEEEmembership{Member,~IEEE,}
        Ping Wang,~\IEEEmembership{Member,~IEEE,}
        Qingjiang Shi,~\IEEEmembership{Senior Member,~IEEE,}
        Xu Zhu,~\IEEEmembership{Senior Member,~IEEE,}\\
        Christos Masouros,~\IEEEmembership{Senior Member,~IEEE}

\thanks{Zhongxiang Wei, Qingjiang Shi, and Ping Wang are with the College of Electronic and Information Engineering, at Tongji University, Shanghai 200092, China. Email: \{z\_wei, shiqj, pwang\}@tongji.edu.cn. }

\thanks{Xu Zhu is with the School of Electronic and Information Engineering, Harbin Institute of Technology, Shenzhen 518055, China. Email:
xuzhu@stu.hit.edu.cn.}
\thanks{Christos Masouros is with the Department of Electronic and Electrical
Engineering, University College London, London WC1E 6BT, U.K. Email:
c.masouros@ucl.ac.uk.}


}

\maketitle

\begin{abstract}

In last decades, dynamic resource programming in partial resource domains has been extensively investigated for single time slot optimizations.
However, with the emerging real-time media applications in fifth-generation communications, their new quality of service requirements are often measured in temporal dimension.
This requires multistage optimization for full resource domain dynamic programming. 
Taking experience rate as a typical temporal multistage metric,  we jointly optimize time, frequency, space and power domains resource for multistage optimization. 
To strike a good tradeoff between system performance and computational complexity, we first transform the formulated mixed integer non-linear constraints into equivalent convex second order cone constraints, by exploiting the coupling effect among the resources. Leveraging the concept of  structural sparsity, the objective of max-min experience rate is given as a weighted 1-norm term associated with the precoding matrix.
Finally, a low-complexity iterative algorithm is proposed for full resource domain programming, aided by another simple conic optimization for obtaining its feasible initial result.
Simulation verifies that our design significantly outperform the benchmarks while maintaining a fast convergence rate,  shedding  light on full  domain dynamic resource programming of  multistage optimizations. 


\end{abstract}


\IEEEpeerreviewmaketitle

\maketitle

\section{Introduction}

Due to the explosive growth of user demands on ubiquitous access and multimedia services, 
dynamic resource  programming has been a fundamental task in the design and management of communication networks in last decades.
The importance of resource programming can be attributed to its key role in the efficient utilization of limited wireless resource as well as interference mitigation, thereby optimizing different system utilities, such as sum rate \cite{Christensen2008Weighted} \cite{Shi2011An} \cite{Bandi2020A}, power consumption \cite{Peel2005A} \cite{Sidiropouls2006Transmit} \cite{Wei2020Multi}, proportional user fairness \cite{Shen2005Adaptive} \cite{Chen2020The}, energy efficiency \cite{Cui2004Energy} \cite{Ngo2013Energy} \cite{Wei2015Full}, latency \cite{Berry2002Communication} \cite{Bettesh2006Optimal}
 \cite{Bai2020L}, security \cite {Khisti2010Secure} \cite{Wei2022Physical}, among others. The resource programming design for optimizing the above utility functions has been investigated for broadcast channel, multiple access  channel, interference channel, and relay channel \cite{Hong2012Signal}. 
In general, the wireless resource to be optimized includes time, frequency, space and power \cite{Castaneda2017An}. For example, allocating orthogonal time or frequency resources to users yields to classic time- and frequency- division multiplex access (FDMA) networks, while joint power and subcarrier allocation of frequency domain can be extensively found in orthogonal frequency division multiple access (OFDMA) networks \cite{Huang2009Joint}. 


With multi-antenna configuration, power and space resources can be well merged into precoding design, such as the well-known water-filling based zero-forcing and minimum mean square error (MMSE) precoders \cite{Peel2005A}. Also, targeting at optimizing different utility functions, optimization-based precoders have been well investigated in last decades. 
As the pioneers, the authors in \cite{Sidiropouls2006Transmit} \cite{Wiesel2006Linear} introduced the use of convex optimization approach for precoding design. In particular,  a power minimization criterion and a max-min SINR fairness criterion were formulated, which can be solved with semi-definite programming (SDP) \cite{Sidiropouls2006Transmit} or conic optimization \cite{Wiesel2006Linear}.
For weighted sum-rate maximization, a more intractable utility function, the authors in \cite{Christensen2008Weighted} \cite{Shi2011An} exploited the hidden relationship between sum-rate and MMSE of decoding data, where some iterative algorithms (referred to as ``weighted MMSE'' approach in follows) were proposed for multiple-input and multiple-output (MIMO) broadcast and interfering broadcast channel. 
In recent years, the optimization based precoding designs have been extended to massive MIMO \cite{Yang2015Fifty}, millimeter wave \cite{Sohrabi2016Hybrid}, hybrid beamforming \cite{Zhu2017A}, and so on.
Further considering frequency domain, it evolves to multi-antenna OFDMA systems \cite{Ng2012Dynamic} \cite{Xiao2015Energy} \cite{Femenias2016Scheduling} \cite{Yu2006Dual}    \cite{Wei2016Energy} \cite{Ng2012Energy}.
Due to the binary decision of subcarrier allocation, resource programming of MIMO OFDMA  always leads to a mixed integer non-linear programming (MINLP) problem.
In \cite{Xiao2015Energy} and \cite{Femenias2016Scheduling}, the authors optimized energy efficiency and sum-rate for MIMO OFDMA systems, where
subcarrier allocation and precoding  are jointly designed based on the Lagrange dual method. 
Although the primary problem is non-convex, it was shown that the duality gap is negligible with a sufficient large number of subcarriers \cite{Yu2006Dual}.
In a similar manner, researchers  investigated  subcarrier allocation and precoding for full duplex relaying systems in \cite{Ng2012Dynamic} \cite{Wei2016Energy}, and for massive-MIMO system in \cite{Ng2012Energy}.
Search based \cite{Lin2009Optimal} \cite{Gotsis2009Dynamic} or binary variable relaxation based approaches \cite{Ng2016Power}  \cite{Liu2021Joint} \cite{Khalili2022Resource} are also commonly used for handling the binary subcarrier allocation in MIMO OFDMA systems. Branch-and-bound \cite{Lin2009Optimal} and heuristic algorithms \cite{Gotsis2009Dynamic} search the possible subcarrier allocation procedure along a selected path for approaching optimum. 
Differently, the relaxation based approaches \cite{Ng2016Power} \cite{Liu2021Joint} \cite{Khalili2022Resource} relax the binary subcarrier allocation indicator into a continuous variable in-between [0,1], and subsequently, a penalty term is often added into the objective function, in order to push the relaxed  continuous variable to its boundary, i.e., 0 or 1. 
In general,  the search based  approaches emphasize near-optimum solution while relaxation based approaches obtain low complexity.



When the number of active users in the systems is large, 
it is necessary to select subsets of users for data delivery in each time slot, referred to as user scheduling or admission control. Evidently, user scheduling involves  resource allocation in time domain, and this is again a binary decision  procedure, i.e., scheduling a user or not in the specific time slot. 
An intuitive way is to decouple the procedures of user scheduling and other domain resource into two steps \cite{Cheng2020Joint} \cite{Yoo2007Multi}, 
i.e., scheduling user  first and then allocating other domains resource to the scheduled users. 
This two-step approach generally obtains low computational complexity at the cost of significant performance degradation. 
To obtain near-optimal solution, the authors in \cite{Li2016Multicell} \cite{Yu2013Multicell} proposed iterative decoupled approaches, where the scheduling and precoding  are optimized in their own iteration based on the feedback of the previous iteration.
Instead of updating the variables separately, the authors in \cite{Bandi2020A} proposed an difference-of-convex based iterative algorithm, where the scheduling and precoding variables are optimized jointly based on the previous updates until convergence.
Different from the iterative approaches,   the users in \cite{Hong2013Joint} \cite{Matskani2008Convex}  relaxed the  precoding and user scheduling problem as a penalized second order cone programming (SOCP) or a SDP problem, targeting at maximizing the number of users that can be served at their subscribed quality-of-service (QoS) or maximizing the signal-to-interference-and-noise ratio (SINR) respectively.
Another line of research tries to  mitigate the binary user scheduling variable from the the original MINLP \cite{Ku2015Joint} \cite{Razaviyayn2014Joint}, where joint user grouping and precoding design was investigated for weighted sum rate maximization in MIMO interfering broadcast channel \cite{Ku2015Joint} and cognitive radios \cite{Razaviyayn2014Joint}. 
As an extension, the authors in \cite{Baligh2014Cross} outlined the combination of ``weighted MMSE'' approach with  higher layers designs, such as user admission, user-base station (BS) association, as well as routing design and data flow control. Note that the terminology  ``user association'' often refers to associating an user to his/her neighboring BSs for communication.

\begin{figure}
	\centering
	\includegraphics[width=5.30 in]{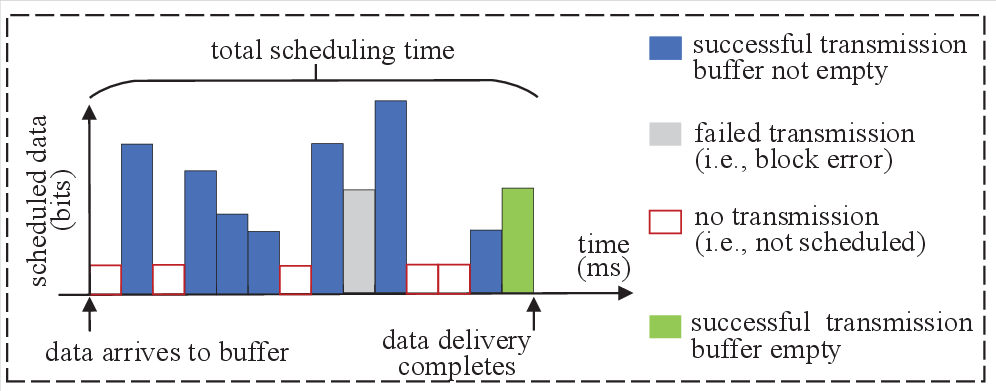}
    \caption{Experience rate is defined as the ratio of a user’s payload size to the total time of complete delivery of its payload, where joint design of  time, frequency, space, and power resources are required.}
    \label{fig:VS1} 
\end{figure}

The above works focus on dynamic resource programming  in the current time slot, which is often referred to as single-stage optimization.
In practice, a user' payload is buffered in the BS for scheduling, and the QoS of many real-time applications is highly dependent to the overall time of complete delivery of the user's payload. 
Recently proposed by 3rd Generation Partnership Project (3GPP),  experience rate is defined as the ratio of a user's payload size to the total time of complete delivery of its payload \cite{3GPP2022}, as demonstrated in Fig. 1. As  each user's delivery time contains the times  of being scheduled and not being scheduled,
failing to schedule the user  with poor channel or large payload early makes it difficult to serve those users in subsequent slots from the perspective of experience rate maximization.
To capture the dynamic resource  programming in temporal dimension, it essentially requires multistage optimization design. 
The authors  in \cite{Bandi2022Joint} investigated overall delivery time minimization, which however only contains user scheduling and precoding design, without the consideration of subcarrier allocation in frequency domain. In addition, the binary variable of user scheduling is relaxed and a  penalty term is added the objective function, which requires a dedicated setup of penalty parameter to avoid significant performance loss. To the best of our knowledge, there is no published research on joint time, frequency, space, and power domains resource programming, where the major difficulties lie in the following aspects. 
\begin{itemize}

\item Both the time slot and subcarrier allocation involve binary decision procedure, and those binary variables are also coupled with the precoding matrix  in the optimization. \item More significantly, 
taking experience rate optimization as an example for the considered multistage optimization, the resource scheduling policy in the current time slot has an impact on the scheduling result in the future  time slot.  
As a result, the coupling effects of those resources not only appear in each slot, but also extend to temporal dimension.
\item The lack of convexity (or more generally, the lack of convex reformulation and transformation) makes it difficult to optimize the multistage utility functions, while maintaining a  reasonable level of complexity. 
\end{itemize}

Motivated by the aforementioned open challenges, in this paper, we present a first attempt to exploit full domains dynamic resource programming for multistage optimization. Our contributions are summarized as follows.
    
\begin{enumerate}
    \item This is the first work investigating multistage optimization that involves the
    full domains of resources, including time, frequency,  space and power. Considering  experience rate as a typical  metric measured in temporal dimension, a novel multistage MINLP optimization is formulated for maximizing the minimal experience rate  among users.
    \item  To handle the complicated multistage MINLP optimization, a judicious routine for convex reformulation is demonstrated. 
    We first show that for such a multistage optimization, the binary decision procedures of frequency and time domains can be well merged into 
    a single binary procedure only in frequency domain, thereby eliminating the coupling of the binary variables in frequency and time domains. Then, departing from the conventional relaxation-based  or  search-based approaches, we exploit the hidden relation between the binary subcarrier allocation variable and its associated precoding matrix, and further remove the binary subcarrier allocation variable without loss of optimality. Further, based on the concept of structural group sparsity, we transform the max-min experience rate objective into a more tractable weighted 1-norm form, associated with the column index of the precoder matrix.  A standard convex reformulation is finally made by transforming the fractional structured SINR as second order cone (SOC) constraints.   
    \item A novel full domain dynamic resource programming (FDRP) algorithm is designed. 
    For obtaining a feasible initial result, a one-shot optimization problem is formulated and solved as a SOCP.     
    With the obtained feasible initial result, the proposed FDRP algorithm updates the variables iteratively and is able to converge to a stationary point after a couple of iterations.  It shows that the whole algorithm design in fact solves a number of conic optimization problems with fast convergence rate, where its complexity is maintained at a low level.
    \item Our study also reveals some important properties of the multistage MINLP design for dynamic full domains resource programming.  As discussed in Remarks 1 and 2, the resources allocated to users are presented in a sparse manner, which in fact facilitates the merging of resource allocation variables for significantly simplifying the algorithm design. Also, by removing the subcarrier allocation variable in P3, it becomes possible to multiplex more users per subcarrier than the number of transmit-antennas $N_t$. However, it turns out that the number of the multiplexed users per subcarrier is generally not larger than $N_t+3$, for maintaining a reasonable value of per-subcarrier SINR.
\end{enumerate}


\textit{Notations}: Matrices and vectors are represented by boldface capital and lower case letters, respectively.
$||\cdot||_p$ denotes $p$-norm. $(\cdot)^H$ denotes hermitian transpose of  a matrix. $|\cdot|$ denotes cardinality of a set or absolute value of a complex variable. $\mathcal{CN}(\cdot)$ denotes complex Gaussian distribution.
Table I summarizes the notations
for the most commonly used variables in this paper.

\begin{table*}
\centering
\caption{Notation Glossary}
\begin{tabular}{ | c | c | c |c|}
      \hline
      Variables & Definitions                 & Variables     & Definitions\\
      \hline
      \hline
      $N$       & Number of subcarriers       & $K$           & Number of users\\
      \hline
      $N_t$     & Number of transmit-antennas & $\bm{h}_{kn}$ & \makecell{Channel vector of user $k$\\ on subcarrier $n$ in the $t$-th slot}\\
      \hline
      $\bm{w}_{knt}$ & \makecell{Precoder of user $k$ on \\subcarrier $n$ in the $t$-th slot} &
      $ \bar{\bm{w}}_{knt}$ & \makecell{Equivalent precoder of user $k$ on \\ subcarrier $n$ in the $t$-th slot}\\
      \hline
      $\bar{\bm{W}}$    & \makecell{Precoder matrix including all users' precoders \\    across $N$ subcarriers in $T$ slots (see \eqref{eq:precoder matirx})} &
      $\bar{\bm{w}}_k$ &  \makecell{Precoder matrix including user $k$'s precoders \\    across $N$ subcarriers in $T$ slots (see \eqref{eq:precoder matirx})}\\
      \hline
      $s_{knt}$    & \makecell{Symbol of user $k$ on \\subcarrier $n$ in the $t$-th slot} &
      $y_{knt}$    &\makecell{Received signal of user $k$\\ on subcarrier $n$ in the $t$-th slot}\\
      \hline
      $z_{knt}$    & Noise of user $k$ on subcarrier $n$ in the $t$-th slot &
      $r_{knt}$    & \makecell{Throughput of user $k$ on subcarrier\\ $n$ in the $t$-th slot, in bits/s}\\
      \hline
      $N_0$ & Noise power spectral density, in dBm/Hz 
      & $B$ & Per-subcarrier bandwidth, in Hz\\
      \hline
      $\Gamma_{knt}$ & \makecell{Receive-SINR of user $k$ on \\ subcarrier $n$ in the $t$-th slot (see \eqref{eq:SINR})} &
      $\bar{\Gamma}_{knt}$ & \makecell{Equivalent SINR of user $k$ \\on subcarrier $n$ in the $t$-th slot (see \eqref{eq:newsinr})}\\
      \hline
      $\alpha_{kt}$ & Binary scheduling variable of time domain &
      $\beta_{knt}$ & Binary scheduling variable of frequency domain\\
      \hline
      $Q_k$ & Payload of user $k$, in bits &
      $T_k$ & \makecell{Overall time slots for completing\\ user $k$'s payload, in s} \\
      \hline
      $T$ & \makecell{Overall time slots for \\ completing all users payload,  in s} &
      $R_k$ & Experience rate of user $k$, in bits/s\\
       \hline
      $\iota$ & Length of each time slot, in ms &
      $P$ & Transmission power budget, in mW\\
      \hline
      $\gamma$ & Optimal result of optimization &
      $\lambda_t$    & \makecell{Weight factor w.r.t    position \\index of  column of matrix (see \eqref{eq:gt})} \\
      \hline
      $o$ & On the order of $o=\mathcal{O}(KNTN_t)$  (see \eqref{eq:compelxity33}) &
      $\epsilon$ & $\epsilon$-optimal accuracy factor (see \eqref{eq:compelxity33})\\
      \hline
      $\zeta_{knt}$ & \makecell{Auxiliary variable for lower bounding the value \\ of   $1+\bar{\Gamma}_{knt}$, i.e., $1+\bar{\Gamma}_{knt} \geq \zeta_{knt}$ (see \eqref{eq:c2ac2b})} &
      $\bm{\zeta}$ & Auxiliary matrix including all $\zeta_{knt}$ (see \eqref{eq:zeta matirx})\\
      \hline
      $a_{knt}$ & Auxiliary variable related to SINR (see \eqref{eq:DC}) &
      $b_{knt}$ & Auxiliary variable related to SINR (see \eqref{eq:DC}) \\
\hline

    \end{tabular}
\end{table*}

\section{System Model and Problem Formulation}

\begin{figure}
	\centering
	\includegraphics[width=5.35 in]{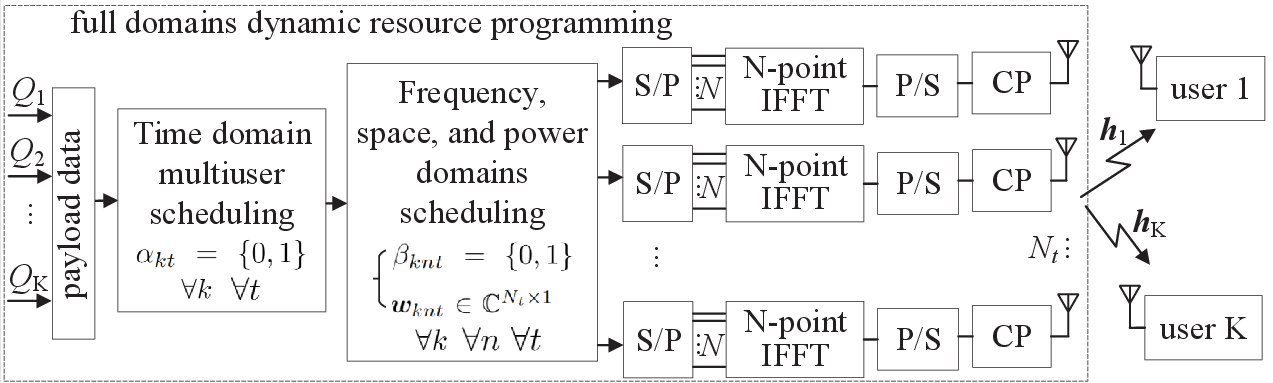}
    \caption{Block diagram for OFDM-based multiuser MIMO communication at downlink. Full domains  resource  are   dynamically scheduled.
    }
    \label{fig:VS1} 
\end{figure}

Consider an OFDM-based MIMO communication at downlink, as depicted in Fig. 2. There are $K$ ($K=|\mathbb{K}|$ with $\mathbb{K}$ denoting user set) users requiring signals from a BS.
$N$-point Inverse Fast Fourier Transform (IFFT) is employed at the BS, and without loss of generality, we assume that all $N$  subcarriers ($N=|\mathbb{N}|$ with $\mathbb{N}$ denoting subcarrier set) are used for data transmission.
The BS is equipped with $N_t$ antennas for transmission, while each user has a single antenna for signal reception.

Define $\bm{w}_{knt} \in \mathbb{C}^{N_t \times 1}$ as the precoding vector employed by the BS, for user $k$ on subcarrier $n$ in the $t$-th slot.
Define $\bm{h}_{knt} \in \mathbb{C}^{ N_t \times 1}$ as the channel from the BS to user $k$ on subcarrier $n$ in the $t$-th slot. Assume that the channels remain unchanged within the coherence time, typically 20 ms (40 time slots) for pedestrian velocity. 
The received signal of user $k$ on subcarrier $n$ in the $t$-th slot is given as 


\begin{equation}
\begin{split}
y_{knt}=\bm{h}^H_{knt} \bm{w}_{knt} s_{knt} + \sum_{k'\ne k}^{K} \bm{h}^H_{knt} \bm{w}_{k'nt} s_{k'nt}+z_{knt},
\label{eq:received signal}
\end{split}
\end{equation}
where $z_{knt} \sim \mathcal{CN}\{0,N_0\}$ denotes receive-noise of user $k$ on subcarrier $n$ in the $t$-th slot, and $N_0$ is the noise power spectral density.
$s_{knt}$ denotes the data symbol for user $k$ on subcarrier $n$ with normalized power.  Define $B$ as the per-subcarrier bandwidth.
It leads to receive-SINR of user $k$ on subcarrier $n$ in the $t$-th slot as

\begin{equation}
\begin{split}
\Gamma_{knt}=\frac{\alpha_{kt} \beta_{knt}|\bm{h}^H_{knt} \bm{w}_{knt}|^2}{\sum_{k'\ne k}^{K} \alpha_{k't} \beta_{k'nt}|\bm{h}^H_{knt} \bm{w}_{k'nt}|^2+BN_0},
\label{eq:SINR}
\end{split}
\end{equation}
where both $\alpha_{kt}=\{0,1\}$ and $\beta_{knt}=\{0,1\}$  are binary variables, $\forall k \in \mathbb{K}$,  $\forall n \in \mathbb{N}$, and $\forall t$. 
$\alpha_{kt}=1$  denotes that user $k$ is scheduled  in the $t$-th time slot, and $\alpha_{kt}=0$ otherwise. $\beta_{knt}=1$ denotes that the user $k$ is allocated with subcarrier $n$ in the $t$-th time slot, and $\beta_{knt}=0$ otherwise.
Hence, the throughput of user $k$ on subcarrier $n$ in the $t$-th slot is calculated as

\begin{equation}
\begin{split}
r_{knt}=B\mathrm{log}_2(1+\frac{\alpha_{kt}\beta_{knt}|\bm{h}^H_{knt} \bm{w}_{knt}|^2}{\sum_{k'\ne k}^{K} \alpha_{k't} \beta_{k'nt}|\bm{h}^H_{knt} \bm{w}_{k'nt}|^2+BN_0}).
\label{eq:throughput}
\end{split}
\end{equation}

Write $Q_k$ as the payload size of user $k$ and $T_k$ as the overall time slots required for complete delivery of user $k$'s payload. Then, the experience rate \cite{3GPP2022} of user $k$ is calculated as

\begin{equation}
\begin{split}
R_k=\frac{Q_k}{T_k},
\label{eq:1}
\end{split}
\end{equation}
where we emphasize that $T_k$ capsules the time of being scheduled, as well as the time of  not being scheduled (in idle). 
Evidently, optimizing $R_k$ involves time, frequency, space and power (space and power  are contained in precoder) domains resource allocation.

Define $\iota$ as the  length of each time slot.
Define $T=\operatorname*{max} \limits_{k \in \mathbb{K} }   \{T_k\}$ as the total number of the required slots for transmitting all $K$ users' data.
Aiming at maximizing the minimum  experience rate among $K$ users, we formulate a multistage full domain dynamic resource  programming, written as

\begin{equation}\boxed{
\begin{split}
&P1~ : \operatorname*{argmax} \limits_{\bm{w}_{knt}, \beta_{knt}, \alpha_{kt}, \forall k \in \mathbb{K}, \forall n \in \mathbb{N}, \forall t }~ \operatorname*{min} ~ \eta_k \frac{Q_k}{T_k  }   ,\\
&\mathrm{s.t.}~(\mathrm{C1}): \sum_{k=1}^{K}\sum_{n=1}^{N} \alpha_{kt} \beta_{knt}||\bm{w}_{knt}  ||_2^2 \leq P,  \forall t =1,...,T,  \\
&~~~~~(\mathrm{C2}): \sum_{t=1}^{T}  \sum_{n=1}^{N} \iota B\mathrm{log}_2(1+\frac{\alpha_{kt}\beta_{knt}|\bm{h}^H_{knt} \bm{w}_{knt}|^2}{\sum_{k'\ne k}^{K} \alpha_{k't}\beta_{k’nt}|\bm{h}^H_{knt} \bm{w}_{k'nt}|^2+BN_0}) \geq Q_k, \forall k \in \mathbb{K},\\
&~~~~~(\mathrm{C3}): \beta_{knt}=\{0,1\}, \forall n \in \mathbb{N}, \forall k \in \mathbb{K}, \forall t =1,...,T,\\
&~~~~~(\mathrm{C4}): \alpha_{kt}=\{0,1\}, \forall k \in \mathbb{K}, \forall t =1,...,T,
\label{eq:p1}
\end{split}}
\end{equation}
where $\eta_k$ denotes the weight factor of user $k$. 
Constraint (C1) denotes transmission power per time slot is bounded by a budget $P$.  Constraint (C2) denotes the overall delivered bits of user $k$ during  $T$ time slots is no smaller than its payload $Q_k$. The binary constraints (C3) and (C4) denote the binary allocation indicators in frequency and time domains.


\section{ Full Domains Resource Multistage Optimization}

P1 is a multistage MINLP problem. Finding its global optimum generally relies on search-based solutions, such as branch-and-bound and knapsack. The complexities of these search-based algorithms increase exponentially with the number of $N$, $K$ and $T$. Thus, in sequel, we focus on the low-complexity algorithm design to find a near-optimum. Algorithm  design is given in subsection III-A,   the approach for obtaining its feasible initial result is given in subsection III-B, and complexity is given in subsection III-C.

\subsection{Algorithm Design}

The first difficulty of solving P1 comes from the combinatorial constraints (C3)-(C4), as well as their coupling with precoder in (C1)-(C2).

\textbf{Remark 1}: User $k$ is said to be scheduled in slot $t$, if at least one subcarrier is allocated to it in the slot. ~~~~~~~~~~~~~~~~~~~~~~~~~~~~~~~~~~~~~~~~~~~~~~~~~~~~~~~~~~~~~~~~~~~~~~~~~~~~~~~~~~~~~~~~~~~~~~$\blacksquare$

Remark 1 is easy to prove. We define a long vector  $[\beta_{k1t}(\bm{w}_{k1t})^H,...,\beta_{kNt}(\bm{w}_{kNt})^H]^H \in \mathbb{C}^{NN_t \times 1}$, $\forall k, t$, which in fact presents  precoder vector for user $k$ across $N$ subcarriers in the $t$-th slot. In particular, if we have

\begin{equation}
\begin{split}
||[\beta_{k1t}(\bm{w}_{k1t})^H,...,\beta_{kNt}(\bm{w}_{kNt})^H]^H||_p
 \neq 0, 
 \label{eq:time scheduling}
\end{split}
\end{equation}
 which means that user $k$ is scheduled in the $t$-th slot and at least one subcarrier is allocated to it, such that
 
\begin{equation}
\begin{split}
&||[\beta_{k1t}(\bm{w}_{k1t})^H,...,\beta_{kNt}(\bm{w}_{kNt})^H]^H||_p\left\{
\begin{array} {lcl}
 \neq 0;~ \mathrm{scheduled ~in~ slot~} t, \\
 = 0 ; ~\mathrm{not~scheduled~ in~ slot~} t.
\end{array} \right.
\label{eq:time scheduling}
\end{split}
\end{equation}

Remark 1 essentially denotes that  merely using the binary variable in frequency domain can well indicate the scheduling policy in time domain. Once the subcarrier allocation policy is decided, 
the binary variable $\alpha_{kt}$  in time domain can be equivalently obtained as

\begin{equation}
\begin{split}
& \alpha_{kt} = \varkappa([\beta_{k1t}(\bm{w}_{k1t})^H,...,\beta_{kNt}(\bm{w}_{kNt})^H]^H), 
\label{eq:relation}
\end{split}
\end{equation}
where the operator $\varkappa(\bm{x})$ returns a value of one if the vector $\bm{x}\neq \bm{0}$ and otherwise. 
Hence, P1 is equivalent to the following optimization

\begin{equation}
\begin{split}
&P2~ : \operatorname*{argmax} \limits_{\bm{w}_{knt}, \beta_{knt}, \forall k \in \mathbb{K}, \forall n \in \mathbb{N}, \forall t }~ \operatorname*{min} ~ \eta_{k} \frac{Q_k}{T_k  }   ,\\
&\mathrm{s.t.}~(\mathrm{C1}): \sum_{k=1}^{K}\sum_{n=1}^{N} \beta_{knt} ||\bm{w}_{knt}  ||_2^2 \leq P,  \forall t =1,...,T,  \\
&~~~~~(\mathrm{C2}):  \sum_{t=1}^{T}  \sum_{n=1}^{N} \iota B\mathrm{log}_2(1+\frac{\beta_{knt}|\bm{h}^H_{knt} \bm{w}_{knt}|^2}{\sum_{k'\ne k}^{K} \beta_{k’nt}|\bm{h}^H_{knt} \bm{w}_{k'nt}|^2+BN_0}) \geq Q_k, \forall k \in \mathbb{K},\\
&~~~~~(\mathrm{C3}): \beta_{knt}=\{0,1\}, \forall n \in \mathbb{N}, \forall k \in \mathbb{K}, \forall t =1,...,T,
\label{eq:p1}
\end{split}
\end{equation}
where the binary variable in time domain has been removed in P2.
However, the coupling between the binary subcarrier allocation variable $\beta_{knt}$ and precoder matrix still hinders algorithm design. Conventional approaches either relax the subcarrier allocation variable into continuous variable within [0,1], or are based on search. These approaches generally 
obtain a solution far from optimum or incur high complexity. In the following, Remark 2 is presented to discuss the sparsity of the resource allocated to users, which is used to reveal the relation between the subcarrier allocation variable and precoder.


\textbf{Remark 2}: The precoders of each user across $N$ subcarriers in $T$ slots are presented in a sparse manner, due to the following reasons:

\begin{itemize}

\item Since user $k$ may only be allocated parts of subcarriers when it is scheduled in the $t$-th slot, the vector $[\beta_{k1t}(\bm{w}_{k1t})^H,...,\beta_{kNt}(\bm{w}_{kNt})^H]^H$  may contain non-zero and  zero elements. In particular, if subcarrier $n$ is allocated to user $k$ in the $t$ slot, i.e., $\beta_{knt}=1$, the associated precoder $\bm{w}_{knt}\ne \bm{0}$, and otherwise.

\item User $k$ may not be consecutively scheduled across $T_k$ slots. If user $k$ is not scheduled in the $t$-th slot, the elements of the vector $[(\bm{w}_{k1t})^H,...,(\bm{w}_{kNt})^H]^H$  are all-zero elements.
 
 \item User $k$ only needs $T_k$ slots for obtaining its complete payload, which is  smaller or equivalent to $T$, i.e., $T_k\leq T$. In this case, the precoders in the last $T-T_k$ slots 
 are all-zero, meaning that user $k$'s data has been delivered and there is no need to schedule it. $~~~~~~~~~~~~~~~~~~~~~~~~~~~\blacksquare$
 
\end{itemize}

Remark 2 in fact reveals the relation between the variable $\beta_{knt}$ and $\bm{w}_{knt}$. More specifically,  the precoder  $\bm{w}_{knt}= \bm{0}$ if and only if (i.i.f) the corresponding frequency domain scheduling variable $\beta_{knt}=0$, otherwise, $\bm{w}_{knt} \neq \bm{0}$  i.i.f $\beta_{knt} = 1$.
Hence, we are able to define an auxiliary variable such that

\begin{equation}
\begin{split}
\bar{\bm{w}}_{knt}=\beta_{knt}\bm{w}_{knt}, \forall k \in \mathbb{K}, \forall n \in \mathbb{N}, \forall t =1,...,T,
\label{eq:bar}
\end{split}
\end{equation}
where the equivalent precoder $\bar{\bm{w}}_{knt}$ relates to the tuple $( \beta_{knt}, \bm{w}_{knt})$ in the form of

\begin{equation}
\begin{split}
& \beta_{knt} = \varkappa(\bar{\bm{w}}_{knt}), \mathrm{and}~ \\
&\beta_{knt} \bm{w}_{knt}=  \bar{\bm{w}}_{knt}.
\label{eq:relation}
\end{split}
\end{equation}

Using $\bar{\bm{w}}_{knt}$ yields an alternative expression of SINR as

\begin{equation}
\begin{split}
\bar{\Gamma}_{knt}=\frac{|\bm{h}^H_{knt} \bar{\bm{w}}_{knt}|^2}{\sum_{k'\ne k}^{K} |\bm{h}_{knt}^H \bar{\bm{w}}_{k'nt}|^2+BN_0}.
\label{eq:newsinr}
\end{split}
\end{equation}

With \eqref{eq:bar}-\eqref{eq:newsinr}, we formulate a new multistage optimization problem as

\begin{equation}
\begin{split}
&P3~ : \operatorname*{argmax} \limits_{\bar{\bm{w}}_{knt}, \forall k \in \mathbb{K}, \forall n \in \mathbb{N}, \forall t }~ \operatorname*{min} ~ \eta_k \frac{ Q_k }{T_k  }    ,\\
&\mathrm{s.t.}~(\mathrm{C1}): \sum_{k=1}^{K}\sum_{n=1}^{N}||\bar{\bm{w}}_{knt}  ||_2^2 \leq P,  \forall t =1,...,T,  \\
&~~~~~(\mathrm{C2}): \sum_{t=1}^{T}  \sum_{n=1}^{N} \iota B \mathrm{log}_2(1+\frac{|\bm{h}^H_{knt} \bar{\bm{w}}_{knt}|^2}{\sum_{k'\ne k}^{K} |\bm{h}^H_{knt} \bar{\bm{w}}_{k'nt}|^2+BN_0}) \geq Q_k, \forall k \in \mathbb{K}.
\label{eq:P1}
\end{split}
\end{equation}

Let tuple $( \beta_{knt}^{\ast}, \bm{w}_{knt}^{\ast}   )$ be the optimal solutions to problem P2, and $ \bar{\bm{w}}_{knt}^{\ast}   $ be the optimal solutions to problem P3. Their  relation is given in Lemma 1.

\textbf{Lemma 1}: Under the transformation of \eqref{eq:relation}, $ \bar{\bm{w}}_{knt}^{\ast}   $ is the optimal solution to P3 i.i.f $( \beta_{knt}^{\ast}, \bm{w}_{knt}^{\ast}   )$ is the optimal solutions to problem P2. ~~~~~~~~~~~~~~~~~~~~~~~~~~~~~~~~~~~~~~~~~~~~~~~~~~~~~~~~~~~~~~~$\blacksquare$

Prove: We first prove its necessity. Let $\bm{\Theta}_2$ and $\bm{\Theta}_3$ denote the feasible sets for P2 and P3. 
Let $\gamma_{\mathrm{(P2)}}$ and $\gamma_{\mathrm{(P3)}}$ denote the optimal values of P2 and P3, respectively.
Assume $\bar{\bm{w}}_{knt}^{\ast}$ is optimal solution to P3. We can construct $( \beta_{knt}^{\ast}, \bm{w}_{knt}^{\ast}   )$ from $\bar{\bm{w}}_{knt}^{\ast}$ by using \eqref{eq:relation}, $\forall k, n,$ and $t$. 
As $(\bm{w}_{knt}^{\ast},\beta_{knt}^{\ast}) \in \bm{\Theta}_2$, it then follows

\begin{equation}
\begin{split}
&\gamma_{\mathrm{(P2)}}(\beta_{knt}^{\ast}, \bm{w}_{knt}^{\ast} )= \gamma_{\mathrm{(P3)}}( \bar{\bm{w}}_{knt}^{\ast} )\\
& ~~~~~~~~~~~~~~~~~~~~\geq  \gamma_{\mathrm{(P3)}}( \bar{\bm{w}}_{knt} ), \forall \bar{\bm{w}}_{knt} \in \bm{\Theta}_3.
\label{eq:AP1}
\end{split}
\end{equation}

For any $( \beta_{knt}, \bm{w}_{knt} ) \in \bm{\Theta}_2$, there exists $\bar{\bm{w}}_{knt} \in \bm{\Theta}_3$ (i.e., let $\bar{\bm{w}}_{knt}= \bm{w}_{knt}$ if $\beta_{knt}=1$, $\forall k, n, t$, and otherwise.  We have

\begin{equation}
\begin{split}
& \gamma_{\mathrm{(P2)}}( \beta_{knt}, \bm{w}_{knt} )= 
 \gamma_{\mathrm{(P3)}} (\bar{\bm{w}}_{knt} ).
\label{eq:AP2}
\end{split}
\end{equation}

From \eqref{eq:AP1} and \eqref{eq:AP2}, it concludes that

\begin{equation}
\begin{split}
& \gamma_{\mathrm{(P2)}} (\beta_{knt}^{\ast}, \bm{w}_{knt}^{\ast} ) \geq  \gamma _{\mathrm{(P2)}}(\beta_{knt}, \bm{w}_{knt} ).
\label{eq:AP3}
\end{split}
\end{equation}

We now  prove its Sufficiency. Similarly, if $(\beta_{knt}^{\ast}, \bm{w}_{knt}^{\ast})$ is the optimal solution of P2, it can be shown from \eqref{eq:relation} that 
$\bar{\bm{w}}_{knt}^{\ast} \in \bm{\Theta}_3$, and

\begin{equation}
\begin{split}
&   \gamma_{\mathrm{(P3)}}(\bar{\bm{w}}_{knt}^{\ast})  = \gamma_{\mathrm{(P2)}} (\beta_{knt}^{\ast}, \bm{w}_{knt}^{\ast}   )  \geq  \gamma_{\mathrm{(P2)}} (\beta_{knt}, \bm{w}_{knt} )  ).
\label{eq:AP4}
\end{split}
\end{equation}

Again, for any $\bar{\bm{w}}_{knt} \in \bm{\Theta}_3$, there exists $(\beta_{knt}, \bm{w}_{knt} )\in \bm{\Theta}_2$ such that

\begin{equation}
\begin{split}
\gamma_{\mathrm{(P3)}}(\bar{\bm{w}}_{knt})  = \gamma_{\mathrm{(P2)}} (\beta_{knt}, \bm{w}_{knt}  ). 
\label{eq:AP5}
\end{split}
\end{equation}

Using \eqref{eq:AP4} and \eqref{eq:AP5}, we have

\begin{equation}
\begin{split}
\gamma_{\mathrm{(P3)}} ( \bar{\bm{w}}_{knt}^{\ast}   )  \geq  \gamma_{\mathrm{(P3)}} ( \bar{\bm{w}}_{knt}   ).
\label{eq:AP6}
\end{split}
\end{equation}

$~~~~~~~~~~~~~~~~~~~~~~~~~~~~~~~~~~~~~~~~~~~~~~~~~~~~~~~~~~~~~~~~~~~~~~~~~~~~~~~~~~~~~~~~~~~~~~~~~~~~~~~~~~~~~~~~~~~~~~~~~~~~~~\square$

Lemma 1 proves that, P3 and P2 shares the same optimal solution. Thus, one can first solve P3  where the binary subcarrier allocation constraint is omitted, and then use \eqref{eq:relation} obtain the optimal tuple $( \beta_{knt}^{\ast}, \bm{w}_{knt}^{\ast}   )$. 
Now, the second difficulty lies in the quadratic-over-quadratic from in (C2). Hence, we introduce an auxiliary variable $\zeta_{knt}$, and transform (C2) in an equivalent form

\begin{equation}
\begin{split}
&(\mathrm{C2a}): \sum_{t=1}^{T}  \sum_{n=1}^{N} \mathrm{log}_2(\zeta_{knt}) \geq \frac{Q_k}{\iota B}, \forall k,\\
& (\mathrm{C2b}):  1+\frac{|\bm{h}^H_{knt} \bar{\bm{w}}_{knt}|^2}{\sum_{k'\ne k}^{K} |\bm{h}^H_{knt} \bar{\bm{w}}_{k'nt}|^2+BN_0} \geq \zeta_{knt}, \forall k, n, t.
\label{eq:c2ac2b}
\end{split}
\end{equation}

The fractional structured expression in (C2b) can be transformed into a difference structure  such that

\begin{equation}
\begin{split}
&\mathrm{(C2b)}:\underbrace{(\sum_{k'=1, k'\ne k}^{K} |\bm{h}^H_{knt} \bar{\bm{w}}_{k'nt}|^2+BN_0)}_{a_{knt}} - \underbrace{\frac{\sum_{k'=1}^{K} |\bm{h}^H_{knt} \bar{\bm{w}}_{knt}|^2+BN_0}{\zeta_{knt}}}_{b_{knt}} \leq 0.
\label{eq:DC}
\end{split}
\end{equation}

The first term of left hand side (i.e., $a_{knt}$) is convex to $\bar{\bm{w}}_{k'nt}$, and the second term of left hand side (i.e., $b_{knt}$) is joint convex to $(\bar{\bm{w}}_{knt}, \zeta_{knt})$, $\forall k, n, t$. 
Hence, \eqref{eq:DC} is a difference between two convex terms. 
By  approximating the term $b_{knt}$ by its first order Taylor expansion, \eqref{eq:DC} can finally be demonstrated into a convex form. Note that this approach generally requires an iteration-based algorithm for obtaining a tight approximation of the relaxed part, which will be detailed later.
For the the sake of notation simplicity, we now define two matrices $\bar{\bm{W}}$ and $\bm{\zeta}$ to include the variables  $\bar{\bm{w}}_{knt}$ and $ \zeta_{knt}$, $\forall k, n, t$,  such that

\begin{equation}
\begin{split}
&\bar{\bm{W}}=\left[\begin{array}{ccccccc}
\bar{\bm{w}}_{111},&...&\bar{\bm{w}}_{11T},&...&\bar{\bm{w}}_{K11}&...&\bar{\bm{w}}_{K1T}\\
&\ddots & &\vdots&&\ddots\\
\bar{\bm{w}}_{1N1},&...&\bar{\bm{w}}_{1NT},&...&\bar{\bm{w}}_{KN1}&...&\bar{\bm{w}}_{KNT} \end{array}\right]\\
&~~~~= \left[\begin{array}{ccc}
\bar{\bm{w}}_{1},&...,&\bar{\bm{w}}_{K}
\end{array}\right]_{NN_t \times KT} ,
\label{eq:precoder matirx}
\end{split}
\end{equation}

and

\begin{equation}
\begin{split}
&\bm{\zeta}=\left[\begin{array}{ccccccc}
\bm{\zeta}_{111},&...&\bm{\zeta}_{11T},&...&\bm{\zeta}_{K11}&...&\bm{\zeta}_{K1T}\\
&\ddots & &\vdots&&\ddots\\
\bm{\zeta}_{1N1},&...&\bm{\zeta}_{1NT},&...&\bm{\zeta}_{KN1}&...&\bm{\zeta}_{KNT} \end{array}\right]_{N \times KT}.
\label{eq:zeta matirx}
\end{split}
\end{equation}

In particular, $\bar{\bm{w}}_{k}$ in \eqref{eq:precoder matirx} is given as

\begin{equation}
\begin{split}
\bar{\bm{w}}_{k}=\left[ \begin{matrix} \bar{\bm{w}}_{k11} & ... &\bar{ \bm{w}}_{k1T}\\ 
 & \ddots& \\  
 \bar{\bm{w}}_{kN1}  & ... & \bar{ \bm{w}}_{kNT}  \end{matrix} \right], \forall k \in \mathbb{K}, 
\label{eq:zeta matirx}
\end{split}
\end{equation}
which in fact represents the precoder of user $k$ across $N$ subcarriers in $T$ slots.

Introduce a superscript $i$ to denote the iteration index.
Let $(\bm{\bar{W}}^{(i)},\bm{\zeta}^{(i)})$ denote  the value of $(\bm{\bar{W}},\bm{\zeta})$ at the  $i$-th iteration.  
Since $b_{knt}(\bm{\bar{W}},\bm{\zeta})$ is convex and differentiable on the considered
domain, one can easily find an
affine majorization as its first order approximation. Hence, we convexify the term $b_{knt}(\bm{\bar{W}},\bm{\zeta})$  
by its affine majorization at a neighborhood of $(\bm{\bar{W}}^{(i)},\bm{\zeta}^{(i)})$, such that

\begin{equation}
\begin{split}
&b_{knt}^{(i)}(\bm{W},\bm{\zeta} ) = b_{knt}(\bm{W}^{(i)},\bm{\zeta}^{(i)} )+\\
& ~~~~~~~~~~~~~~~~~ \mathrm{Re} \bigg( \sum_{k'=1}^{K} (\frac{ \partial b_{k'nt}  }{\partial \bar{\bm{w}}_{k'nt}   })^H ( \bar{\bm{w}}_{k'nt}-\bar{\bm{w}}_{k'nt}^{(i)} ) +    (\frac{ \partial b_{k'nt}  }{\partial \zeta_{knt}   })^H ( \zeta_{k'nt}-\zeta_{k'nt}^{(i)} )      \bigg), 
\label{eq:P1}
\end{split}
\end{equation}
which is further given as

\begin{equation}
\begin{split}
&b_{knt}^{(i)}(\bm{W},\bm{\zeta} )  = b_{knt}(\bm{W}^{(i)},\bm{\zeta}^{(i)} )+ \mathrm{Re} \bigg(
\sum_{k'=1}^{K}  \big(\frac{ 2\bm{h}_{knt} \bm{h}_{knt}^H \bar{\bm{w}}_{k'nt}^{(i)}}{\zeta_{knt}^{(i)}} \big)^H (\bar{\bm{w}}_{k'nt}-\bar{\bm{w}}_{k'nt}^{(i)}   )  -\\
&~~~~~~~~~~~~~~~~~~~~~~~~~~~~~~~~~~~~~~~~~~~~~ \big(\frac{ \sum_{k'=1}^{K} |\bm{h}^H_{kn} \bar{\bm{w}}_{k'nt}^{(i)}|^2+BN_0    }{ (\zeta_{knt}^{(i)})^2  } \big)^H ( \zeta_{knt}-\zeta_{knt}^{(i)}  ) \bigg).
\label{eq:firstorder}
\end{split}
\end{equation}

Substituting \eqref{eq:firstorder} into \eqref{eq:DC}, we can re-formulate (C2b) as

\begin{equation}
\begin{split}
& (\tilde{\mathrm{C2b}}): (\sum_{k'=1, k'\ne k}^{K} |\bm{h}^H_{knt} \bar{\bm{w}}_{k'nt}|^2+BN_0)-b_{knt}(\bar{\bm{W}}^{(i)},\bm{\zeta}^{(i)} )-\\
&~~~~~~~~~~~\mathrm{Re} \bigg(
\sum_{k'=1}^{K}  \big(\frac{ 2\bm{h}_{knt} \bm{h}_{knt}^H \bar{\bm{w}}_{k'nt}^{(i)}}{\zeta_{knt}^{(i)}} \big)^H (\bar{\bm{w}}_{k'nt}-\bar{\bm{w}}_{k'nt}^{(i)}   )  -\\
&~~~~~~~~~~~~~~~~ \big(\frac{ \sum_{k'=1}^{K} |\bm{h}^H_{knt} \bar{\bm{w}}_{k'nt}^{(i)}|^2+BN_0   }{ (\zeta_{knt}^{(i)})^2  } \big)^H ( \zeta_{knt}-\zeta_{knt}^{(i)}  ) \bigg) \leq 0, \forall k, n, t,
\label{eq:P1}
\end{split}
\end{equation}
which becomes a convex constraint with respect to (w.r.t) variables. It is important to note that the convexification step provides an upper bound for the original non-convex difference-of-convex constraint (C2b). Hence, $(\mathrm{C2b})$ always holds when $(\tilde{\mathrm{C2b}})$ holds. 

Now, we turn to handling the objective function. 
As both $Q_k$ and $T_k$ are non-negative variables, the max-min objective function can be optimized equivalently  as a  min-max function, written in the form of

\begin{equation}
\begin{split}
&P4~ : \operatorname*{argmin} \limits_{\bar{\bm{W}}, \bm{\zeta} }~ \operatorname*{max} ~ \{ \frac{T_k}{ \eta_k Q_k } \}  ,\\
&\mathrm{s.t.}~(\mathrm{C1}), (\mathrm{C2a}), (\tilde{\mathrm{C2b}}).
\label{eq:P1}
\end{split}
\end{equation}


The  last difficulty comes from finding a proper formulation to calculate $T_k$ in the objective function.
Remark 2 reveals that, there is no resource allocated to user $k$ in the remaining $T-T_k$ slots. 
By \eqref{eq:precoder matirx},  the  highest position index of the non-zero column of  $\bar{\bm{w}}_{k}$ (i.e., the value of $T_k$)  directly measures the transmission time for user $k$. An operator $g(\bm{V})$ that returns the highest position index of the non-zero column  of a matrix $\bm{V}$ would be the optimal formulation for calculating $T_k$, written as

\begin{equation}
\begin{split}
g(\bm{V})=\left\{
\begin{array}{rcl}
\mathrm{max} \{t, || \bm{v}_t ||_1   \ne 0  \},   &  \bm{V} \ne \bm{0}, \\
0~~~~~~~~~~~~,   &  \bm{V}=\bm{0},
\end{array} \right .
\label{eq:optimalTk}
\end{split}
\end{equation}
where $\bm{v}_t$ is the $t$-th column of the matrix $\bm{V}$.
Nevertheless,  the formulation in \eqref{eq:optimalTk} is not a convex function.
Hence, we alternatively provide a tractable expression, which equivalently optimizes the experience rate among users. 
Intuitively, $g(\cdot)$ should be a non-decreasing function w.r.t the position index of the non-zero columns.
When minimizing the value of $g(\cdot)$, 
it lets $\bar{\bm{w}}_k$ tend to have non-zero elements located in the leftmost columns, but to have zero-elements located in the rightmost columns.
As a result, it makes the highest position index  of the  non-zero column of $\bar{\bm{w}}_k$  small, thereby equivalently yielding a small value of $T_k$.
Motivated by the concept of structural group sparsity \cite{Bach2012Structured}, 
we therefore formulate $g(\cdot)$ as 


\begin{equation}
\begin{split}
g(\bm{V})= \sum_{t=1}^{T}\lambda_t  || \bm{v}_t  ||_1, 
\label{eq:gt}
\end{split}
\end{equation}
where $\lambda_t$ is a non-decreasing weighted factor w.r.t. position index of columns, i.e., the value of $t$. 
For measuring the structural group sparsity of user $k$'s precoder matrix $\bar{\bm{w}}_k$, its $t$-th column is exactly the user's precoders across $N$ subcarriers in the $t$-th slot, i.e.,
$[ \bar{\bm{w}}_{k1t}^H,...,\bar{\bm{w}}_{kNt}^H  ]^H$.
Hence, aided by the formulation of \eqref{eq:gt}, we measure the structural group sparsity of user $k$ as

\begin{equation}
\begin{split}
g(\bar{\bm{w}}_k)= \sum_{t=1}^{T}\lambda_t  || [ \bar{\bm{w}}_{k1t}^H,...,\bar{\bm{w}}_{kNt}^H  ]^H  ||_1, 
\label{eq:gt6}
\end{split}
\end{equation}
which leads to the following optimization problem



\begin{equation}
\begin{split}
&P5~ : \operatorname*{argmin} \limits_{\bar{\bm{W}},\bm{\zeta} }~ \operatorname*{max} ~ \{\frac{\sum_{t=1}^{T}\lambda_t||[ \bar{\bm{w}}_{k1t}^H,...,\bar{\bm{w}}_{kNt}^H  ]^H||_1 }{ \eta_k Q_k } \}  ,\\
&\mathrm{s.t.}~(\mathrm{C1}), (\mathrm{C2a}), (\tilde{\mathrm{C2b}}).
\label{eq:P4}
\end{split}
\end{equation}

Evidently, optimizing the objective in \eqref{eq:P4} is equivalent to optimizing the experience rate among  users.
Let  $\gamma$ serve as the maximum value of 
$\frac{\sum_{t=1}^{T}\lambda_t||[ \bm{w}_{k1t}^H,...,\bm{w}_{kNt}^H  ]^H||_1 
}{ \eta_k Q_k }$ among $K$ users, i,e., 

\begin{equation}
\begin{split}
  \gamma \geq \frac{\sum_{t=1}^{T}\lambda_t||[ \bm{w}_{k1t}^H,...,\bm{w}_{kNt}^H  ]^H||_1 
}{ \eta_k Q_k }   , \forall k \in \mathbb{K},
\label{eq:P4}
\end{split}
\end{equation}

Then, P5  can be transformed into



\begin{equation}\boxed{
\begin{split}
&P6~ : \operatorname*{argmin} \limits_{\bar{\bm{W}},\bm{\zeta} }~  \gamma     ,\\
&\mathrm{s.t.}~(\mathrm{C1}): \sum_{k=1}^{K}\sum_{n=1}^{N}||\bar{\bm{w}}_{knt}  ||_2^2 \leq P,  \forall t =1,...,T,  \\
&~~~~~(\mathrm{C2a}): \sum_{t=1}^{T}  \sum_{n=1}^{N} \mathrm{log}_2(\zeta_{knt}) \geq \frac{Q_k}{\iota B}, \forall k \in \mathbb{K},\\
& ~~~~~(\mathrm{\tilde{C2b}}):  (\sum_{k'=1,k'\ne k}^{K} |\bm{h}^H_{knt} \bar{\bm{w}}_{k'nt}|^2+BN_0)- b_{knt}(\bar{\bm{W}}^{(i)},\bm{\zeta}^{(i)} )-\\
&~~~~~~~~~~~~~~~~ \mathrm{Re} \bigg(
\sum_{k'=1}^{K}  (\frac{ 2\bm{h}_{knt} \bm{h}_{knt}^H \bar{\bm{w}}_{k'nt}^{(i)}}{\zeta_{knt}^{(i)}})^H (\bar{\bm{w}}_{k'nt}-\bar{\bm{w}}_{k'nt}^{(i)}   )  -\\
&~~~~~~~~~~~~~~~~~~~~~~ (\frac{ \sum_{k'=1}^{K} |\bm{h}^H_{knt} \bar{\bm{w}}_{k'nt}^{(i)}|^2+ BN_0   }{ (\zeta_{knt}^{(i)})^2  })^H ( \zeta_{knt}-\zeta_{knt}^{(i)}  ) \bigg) \leq 0, \forall k, n, t\\
&~~~~~(\mathrm{C3}) :\frac{\sum_{t=1}^{T}\lambda_t||[ \bm{w}_{k1t}^H,...,\bm{w}_{kNt}^H  ]^H||_1 
}{ \eta_k Q_k }  \leq \gamma, \forall k \in \mathbb{K},
\label{eq:P1}
\end{split}}
\end{equation}
which becomes a standard convex problem, and is readily solved by CVX. To guarantee performance of P6, it is important to find a feasible initial result $(\bar{\bm{W}}^{(i)},\bm{\zeta}^{(i)})$, as detailed in the next subsection.

\subsection{Feasible Initial Result}

The performance of P6 depends on the choice of the feasible initial result. More importantly, the feasible initial result should locate in the confined domain by those constraints, where the major difficulty lies in the complete delivery constraint in (C2a) and $(\mathrm{\tilde{C2b}})$ in \eqref{eq:c2ac2b}. Hence, we are motivated to propose the following algorithm for finding a feasible initial result.

Let us initialize a positive value of $T$, and assume all users payload can be delivered within $T$ slots.  
For user $k$ with payload $Q_k$, let its data size delivered per-subcarrier in each slot be $\frac{Q_k}{NT}$, and thus its target per-subcarrier SINR is calculated as 

\begin{equation}
\begin{split}
\bar{\Gamma}_{knt}=2^{\frac{Q_k}{NTB\tau}}-1.
\label{eq:FIP11}
\end{split}
\end{equation}

Then we set a corresponding per-subcarrier SINR constraint as

\begin{equation}
\begin{split}
\frac{|\bm{h}^H_{knt} \bar{\bm{w}}_{knt}|^2}{\sum_{k'\ne k}^{K} |\bm{h}_{knt}^H \bar{\bm{w}}_{k'nt}|^2+BN_0} \geq 2^{\frac{Q_k}{NTB\tau}}
-1, \forall k, n, t,
\label{eq:FIP11}
\end{split}
\end{equation}
which can be formulated as a convex SOC constraint

\begin{equation}
\begin{split}
&|| [BN_0, \bm{h}_{knt}^H \bar{\bm{w}}_{1nt},...,\bm{h}_{knt}^H \bar{\bm{w}}_{(k-1)nt},\bm{h}_{knt}^H \bar{\bm{w}}_{(k+1)nt},\\
&~~~~~~~~~~~~~~~~~~~~~...,\bm{h}_{knt}^H \bar{\bm{w}}_{Knt} ]   ||_2 \leq \frac{ \mathrm{Re} \{ \bm{h}_{knt}^H \bar{\bm{w}}_{knt} \} }{\sqrt{2^{\frac{Q_k}{NTB\tau}}
-1} }, \forall k, n, t.
\label{eq:FIP22}
\end{split}
\end{equation}

This constraint helps guarantee the complete delivery of each user's payload, thus well replacing the constraints (C2a) and $(\mathrm{\tilde{C2b}})$. Then we are ready to solve  the following optimization problem

\begin{equation}
\begin{split}
&P7~ : \operatorname*{min} \limits_{\bar{\bm{w}}_{knt}, \forall k, n, t }~   0    ,\\
&\mathrm{s.t.}~(\mathrm{C4}): \sum_{k=1}^{K}\sum_{n=1}^{N}||\bar{\bm{w}}_{knt}  ||_2^2 \leq P,  \forall t,  \\
&~~~~~(\mathrm{C5}): || [BN_0, \bm{h}_{knt}^H \bar{\bm{w}}_{1nt},...,\bm{h}_{knt}^H \bar{\bm{w}}_{(k-1)nt},\bm{h}_{knt}^H \bar{\bm{w}}_{(k+1)nt},\\
&~~~~~~~~~~~~~~~~~~~~~~~~~~~~~~~~~~...,\bm{h}_{knt}^H \bar{\bm{w}}_{Knt} ]   ||_2 \leq \frac{ \mathrm{Re} \{ \bm{h}_{knt}^H \bar{\bm{w}}_{knt} \} }{\sqrt{2^{\frac{Q_k}{NTB\tau}}
-1} },\forall k, n, t,
\label{eq:P5}
\end{split}
\end{equation}
which is a convex SOCP and can be readily solved. 
Importantly, if the feasible set confined by the constraints is empty, one can properly increase the value of $T$, until there is a feasible solution for P7. Then, the obtained result is used as the initial result of P6.
Now, we are able to outline the whole algorithm in the following  Algorithm 1.

\begin{algorithm}
\begin{small}
\caption{The FDRP Algorithm}
\label{alg:Algorithm3}
\begin{algorithmic}[1]
\renewcommand{\algorithmicrequire}{ \textbf{Input:}} 
\renewcommand{\algorithmicensure}{ \textbf{Output:}} 
\REQUIRE  CSI,  power budget $P$, and payload $Q_k$, $\forall k\in \mathbb{K}$.\\
 \textbf{Initialization}:
\STATE Initialize $i=0$. 
\STATE Solve optimization P7 for finding a feasible initial result.
\STATE Obtain the initial result $(\bar{\bm{W}}^{(0)},\bm{\zeta}^{(0)})$, such that
$\bar{\bm{w}}_{knt}^{(0)}=\bar{\bm{w}}_{knt}$, and $ \zeta_{knt} ^{(0)}=1+\frac{|\bm{h}^H_{knt} \bar{\bm{w}}_{knt}|^2}{\sum_{k'\ne k}^{K} |\bm{h}_{knt}^H \bar{\bm{w}}_{k'nt}|^2+BN_0}  $, $\forall k, n, t$.\\
\textbf{Iterative procedure}:
\REPEAT
\STATE Solve optimization P6 to obtain the optimal value $(\bar{\bm{W}}^{\ast},\bm{\zeta}^{\ast})$
\STATE Update $i=i+1$.
\STATE Update $(\bar{\bm{W}^{(i)}},\bm{\zeta}^{(i)})$, such that 
$\bar{\bm{w}}_{knt}^{(i)}=\bar{\bm{w}}_{knt}^{\ast}$, and $ \zeta_{knt} ^{(i)}=1+\frac{|\bm{h}^H_{knt} \bar{\bm{w}}_{knt}^{\ast}|^2}{\sum_{k'\ne k}^{K} |\bm{h}_{knt}^H \bar{\bm{w}}_{k'nt}^{\ast}|^2+BN_0}  $, $\forall k, n, t$.
\UNTIL{Convergence}
\ENSURE Optimal result $(\bar{\bm{W}}^{\ast},\bm{\zeta}^{\ast})$.
\end{algorithmic}
\end{small}
\end{algorithm}

\textbf{Remark 3}: 
Since  the subcarrier allocation variable is omitted, it becomes possible to multiplex more users per subcarrier than the number of transmit-antennas $N_t$. 
By contrast, the previous work \cite{Bandi2020A} applied the relaxation based approach for handling the binary subcarrier allocation variable, where the number of the multiplexed users per subcarrier is strictly no larger than the number of transmit-antennas $N_t$. Though the work in \cite{Ku2015Joint} introduced a semi-definite relaxation (SDR)-based approach for handling  the binary variable, it still tends to multiplex less users than the number of transmit-antennas for obtaining a rank-1 solution. As a result, from the perspective of resource programming, our design is endorsed with higher degrees-of-freedom.   $~~~~~~~~~~~~\blacksquare$

Remark 3 states that the constraint on the number of the multiplexed users per subcarrier can be canceled. However, in practice, the number of users multiplexed on each subcarrier can not be arbitrarily large, as summarized in Remark 4.

\textbf{Remark 4}: 
In general, the number of multiplexed users per subcarrier is  not larger than $N_t+3$, for maintaining a reasonable value of SINR. 
$~~~~~~~~~~~~~~~~~~~~~~~~~~~~~~~~~~~~~~~~~~~~~~~~~~~~~~~~\blacksquare$

Further discussion of Remarks 3-4 will be demonstrated in simulation of Section IV.

\subsection{Complexity Analysis}

Now we analyze the complexity of the proposed algorithm. It first solves optimization P7 for obtaining an initial result, and then iteratively solve optimization P6 with the updated feasible results until convergence. 
Optimization P7 is a SOCP. There are $T$ SOC constraints in (C4) of dimension $KNN_t$, and (C5) can be equivalently decomposed into $KNT$ SOC constraints of (C5a) and $KNT$ linear constraints of (C5b), such as

\begin{equation}
\begin{split}
&(\mathrm{C5a}): || [BN_0, \bm{h}_{knt}^H \bar{\bm{w}}_{1nt},...,\bm{h}_{knt}^H \bar{\bm{w}}_{(k-1)nt},\bm{h}_{knt}^H \bar{\bm{w}}_{(k+1)nt},\\
&~~~~~~~~~~~~~~~~~~~~~~~~~~~~~~...,\bm{h}_{knt}^H \bar{\bm{w}}_{Knt} ]   ||_2 \leq \theta_{knt},\forall k, n, t,\\
&(\mathrm{C5b}): \theta_{knt} \leq \frac{ \mathrm{Re} \{ \bm{h}_{knt}^H \bar{\bm{w}}_{knt} \} }{\sqrt{2^{\frac{Q_k}{NTB\tau}}-1} },\forall k, n, t,
\label{eq:P5}
\end{split}
\end{equation}
where $\theta_{knt}$ is an auxiliary variable. Hence, constraint (C5) equivalently lets P6 subject to $KNT$ SOC constraints of dimension $(K-1)N_t+1$ and $KNT$ linear constraints. 
Given an accuracy factor $\epsilon$,  the computational complexity of solving P7 is calculated as

\begin{equation}
\begin{split}
&C_{\mathrm{P7}}=\mathrm{ln}\frac{1}{\epsilon}
\sqrt{ \underbrace{ 2T+3KNT }_{C_{\mathrm{bar}}} } \cdot \\
&~~~~~~~~\bigg( \underbrace{oT+o^2T+oT(KNN_t)^2+oKNT\big((K-1)N_t+1 \big)^2}_{C_{\mathrm{form}}}   + \underbrace{o^3}_{C_{\mathrm{factor}}}  \big), 
\label{eq:compelxity33}
\end{split}
\end{equation}
where $o$ is on the order of $o=\mathcal{O}(KNTN_t)$. 
Specifically, the term $C_{\mathrm{bar}}$  denotes the so-called barrier parameter, measuring the geometric complexity of the conic constraints of the optimization problem. $C_{\mathrm{form}}$ and $C_{\mathrm{factor}}$ represent the complexities of forming and factorization of a $o \times o$ matrix, which is built to guide the search direction of the interior-point method \cite{Wei2020Multi} \cite{Wang2014Outage}. 
On the other hand, the SOCP P6 is subject to $T$ SOC constraints in (C1) with dimension $KNN_t$, $K$ linear constraints in (C2a), $KNT$ linear constraints in  ($\hat{\mathrm{C2b}}$), and $K$ linear constraints in (C3). 
Hence, the per-iteration computational complexity in P6 is calculated as


\begin{equation}
\begin{split}
&C_{\mathrm{P6}}=\mathrm{ln}\frac{1}{\epsilon} \sqrt{2K+KNT+2T} \cdot\\
&~~~~~~~~\bigg( o(2K+KNT)+o^2(KNT+2K)+oT(KNN_t)^2+o^3     \bigg).
\label{eq:compelxity34}
\end{split}
\end{equation}

Now, the overall complexity of the proposed FDRP algorithm is given as $C_{\mathrm{t}}=C_{\mathrm{P7}}+lC_{\mathrm{P6}}$, where $l$ denotes the iteration numbers for convergence.  In general, P6 converges after a couple of iterations, which will be further shown in simulation part.

\section{Numerical Results}

Our results employ Monte Carlo simulations of the considered problem. 
Unless otherwise specified, the parameters are set as follows. 
The number of subcarriers is set to $N=4$, and per-subcarrier bandwidth is set to 30 kHz. 
The length of time slot is set to $\iota=0.5$ ms. 
The noise spectrum density is $N_0=-174$ dBm/Hz \cite{Ng2012Energy} \cite{Cheng2020Joint}.  Rayleigh fading is adopted for modeling channels. 
The number of transmit-antennas is set to $N_t=5$, and there are $K=6$ users. 
Without loss of generality, we set $\eta_k=1$ for all users.


The following schemes are selected as
benchmarks. 
a) Uniform resource programming (U-RP) based design, where resources of different domains are allocated to users evenly. In fact, this uniform resource programming approach is exactly shown by the optimization problem P7. 
b) Greedy resource programming (G-RP), which allocates all resources to one user and complete the delivery of each user's payload sequentially. In this case, the spatial domain becomes single user multi-output and single-input transmission, where maximal ratio transmit precoding is optimal \cite{Peel2005A}.
c) Semi-definite relaxation based maximum sum-rate resource programming (SDR-RP) \cite{Ku2015Joint}. Towards sum rate maximization in each time slot, the allocation of frequency, power, and space resources are formulated as a SDP problem, until all users' payloads are delivered \footnote{The SDR-RP algorithm in \cite{Ku2015Joint} was designed originally for multi cell cognitive radios. Hence, for a fair comparison, the inter-cell interference constraint between the primary and secondary cognitive radios is canceled.
If the number of multiplexed users per subcarrier is not larger than the number of transmit-antennas, the optimal precoding matrices by the SDR-RP are with rank-1 in most cases \cite{Luo2010Semi}. Then the precoding vector for each user can be obtained by eigenvalue decomposition, and then the transmit-side equalization is applied to align the phase of the precoding vector to the associated channel \cite{Wei2022Fundamentals}.}. 
d) Relaxation based maximum sum-rate resource allocation (R-RP) \cite{Bandi2020A}. Different from the SDR-RP, R-RP relaxes the binary subcarrier allocation procedure into a continuous variable, and then a penalty term is added onto its objective function to push the relaxed variable to [0,1], until all users' payloads are delivered \footnote{The value of the penalty factor is set to 100. In practice, the penalty factor does not necessarily let all the subcarrier allocation variables equal to 0 or 1. Thus, the relaxed subcarrier allocation variables within [0,1] can be treated as  time sharing factors among users on the particular subcarriers. }. Note that subcarrier allocation in frequency domain is not involved in SDR-RP and R-RP algorithms. Hence, their precoders are designed on a subcarrier-by-subcarrier basis.


\begin{figure}
	\centering
	\includegraphics[width=4.5 in]{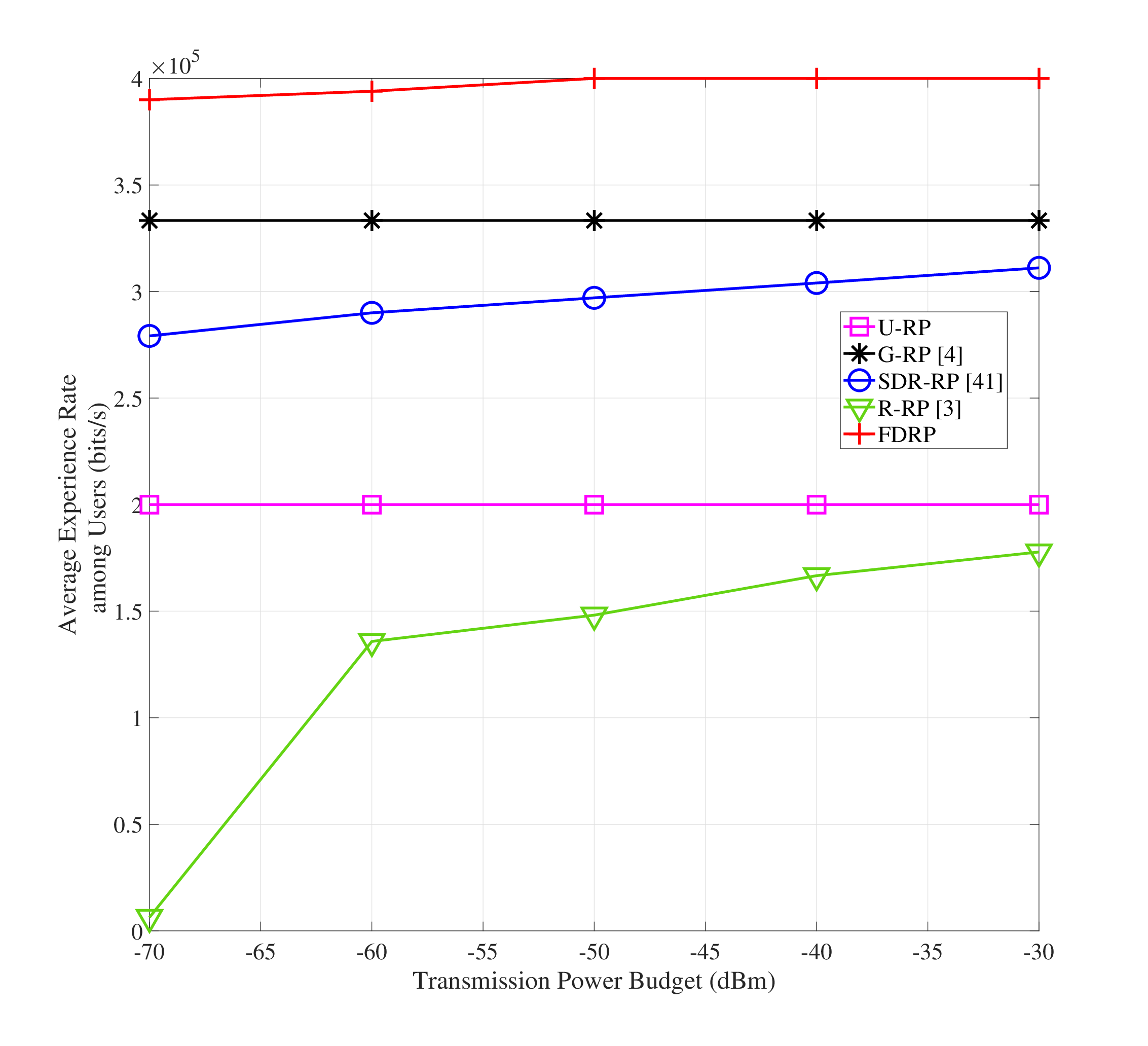}
    \caption{The impact of transmission power budget $P$ on average experience rate among users. Payload is set to $Q_k= 200$ bits for each user, $\forall k \in \mathbb{K}$. The weight  factors in \eqref{eq:gt} are set to $\lambda_1=0.01$ and $\lambda_2=10$ for the FDRP algorithm.
    }
    \label{fig:PvsER} 
\end{figure}

In Fig. \ref{fig:PvsER}, the impact of transmission power budget $P$ on average experience rate among users is demonstrated. 
First, it is observed that the proposed FDRP algorithm obtains the highest level of experience rate, at all transmission power budgets. With -50 dBm or higher transmission power, the FDRP algorithm achieves the highest experience rate, i.e., $4\times10^5$ bits/s, where the delivery of all user's payloads can be completed in the first time slot.
It is because the formulated multistage optimization enables joint design of time, frequency, space, and power resources, thus providing a high DoF for dynamic resource programming. Also, for users with decaying channel and/or large payload, it tends to allocate more resources so that the user's payload can be delivered within a short time. 
In comparisons, two sum-rate maximization algorithms, i.e., SDR-RP and R-RP, aim at maximizing sum-rate in each time slot, and thus always allocate a large portion of resources to the users having good channel quality. As a result, the users having decaying channel or large payload need more time for obtaining complete payload, yielding poor experience rate performance. 
Second, for the purpose of user fairness, the R-RP algorithm sets a subscribed SINR requirement on the scheduled users, which may lead to a lower value of sum-rate  than the SRR-RP algorithm. Also, the number of the served users of the R-RP algorihm  is strictly limited by the number of transmit-antennas. 
Due to the above reasons, the SDR-RP shows a better experience rate performance over the R-RP under the parameters setup of Fig. \ref{fig:PvsER}. 
For the G-RP algorithm, its performance remains unchanged at the demonstrated power range. It is because users are served sequentially by the G-RP algorithm. With adequate resources (i.e., high power budget), each user needs almost fixed number of subcarriers for completing its overall payload, though the multiplexing gain per subcarrier is poor by the G-RP algorithm. 
Third, it demonstrates that a high level of power budget increases the value of experience rate, except of the U-RP algorithm.


\begin{figure}
	\centering
	\includegraphics[width=4.5 in]{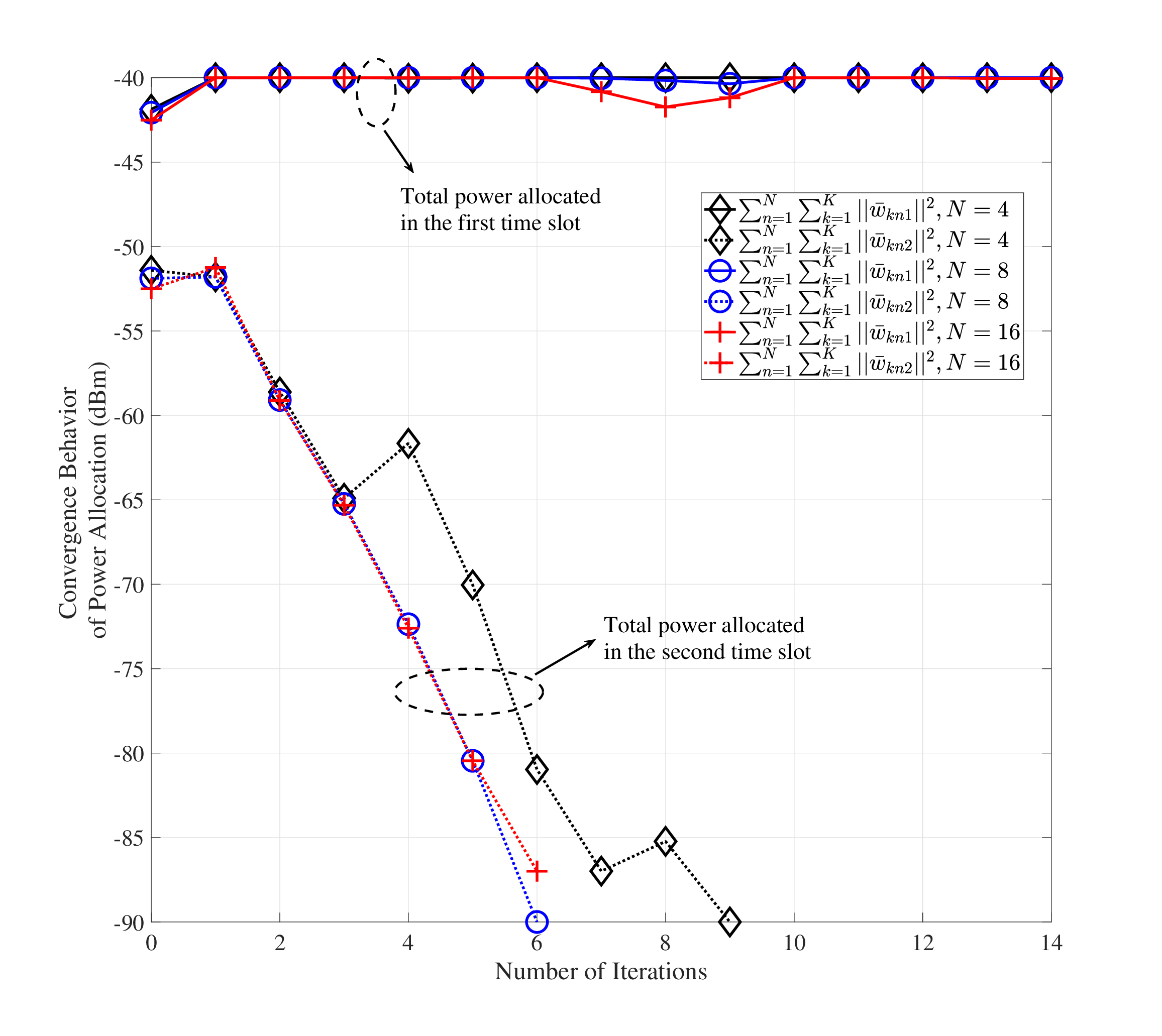}
    \caption{Convergence behavior of the transmission power. Payload is set to $Q_k= 200$ bits for each user, $\forall k \in \mathbb{K}$. Power budget is set to $P=-40$ dBm. The number of subcarriers is set to $N=4, 8$, and $16$.
    }
    \label{fig:convergence} 
\end{figure}

Fig. \ref{fig:convergence} shows the convergence behavior of the transmission power in different time slots. It is observed that the proposed algorithm is able to converge after a few iterations, typically approaching a stationary point within 10 iterations. As the algorithm only needs to solve a conic optimization (i.e., P6) in each iteration, this proves that the whole computational complexity of the proposed algorithm is maintained at a low level. 
Also, it can be seen that transmission power in the first time slot is fully utilized,  while the transmission power of the second slot approaches 0 after 9 iterations. It again means that, the proposed design provides a high level of experience rate for users, and the delivery of all users' payloads can be completed in the first slot. 

\begin{figure}
	\centering
	\includegraphics[width=4.5 in]{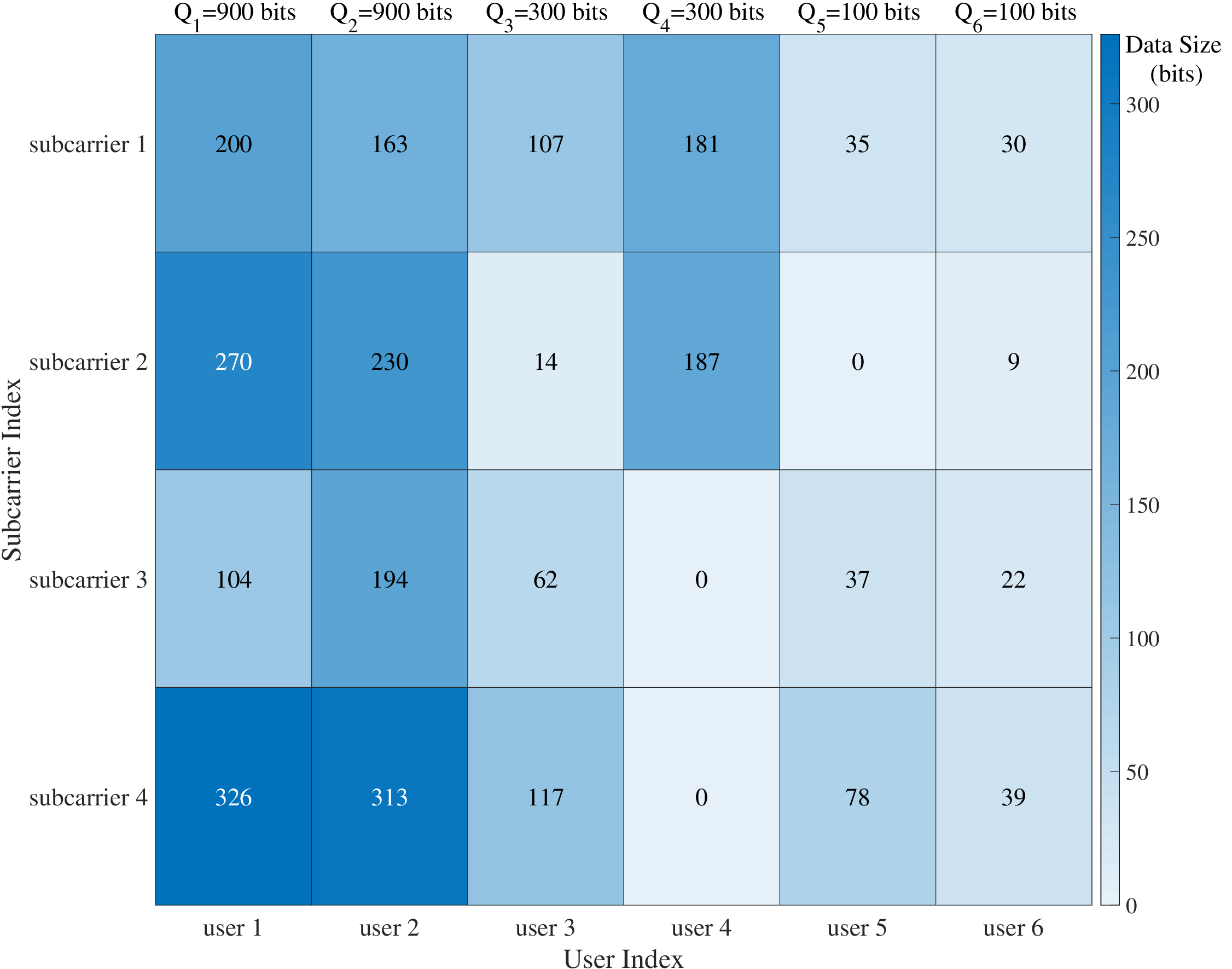}
    \caption{A heat-map of data transmitted of users across different subcarriers, for a specific channel realization.  Payload is set to $[900,900,300,300,100,100]$ bits. Power budget is set to $P=0$ dBm. 
    }
    \label{fig:heatmap} 
\end{figure}

Fig. \ref{fig:heatmap} demonstrates the resource scheduling behavior when users have different payload sizes. It is observed that for the users have large payload sizes (such as users 1 and 2) the proposed FDRP algorithm tends to schedule more resources to them, so that those users receive more data per subcarrier and obtain complete payload soon. 
In comparisons, the algorithm schedules less resources to  the users having small payload size, as it is beneficial to the system experience rate performance.
Also, for the users having identical payload size (such as users 1 and 2, users 3 and 4, users 5 and 6), the resource scheduled in each subcarrier differs.
Based on the observation above, it verifies that both the payload size and  CSI   are captured by the proposed algorithm, for maximizing the experience rate among users.

\begin{figure}
	\centering
	\includegraphics[width=4.5 in]{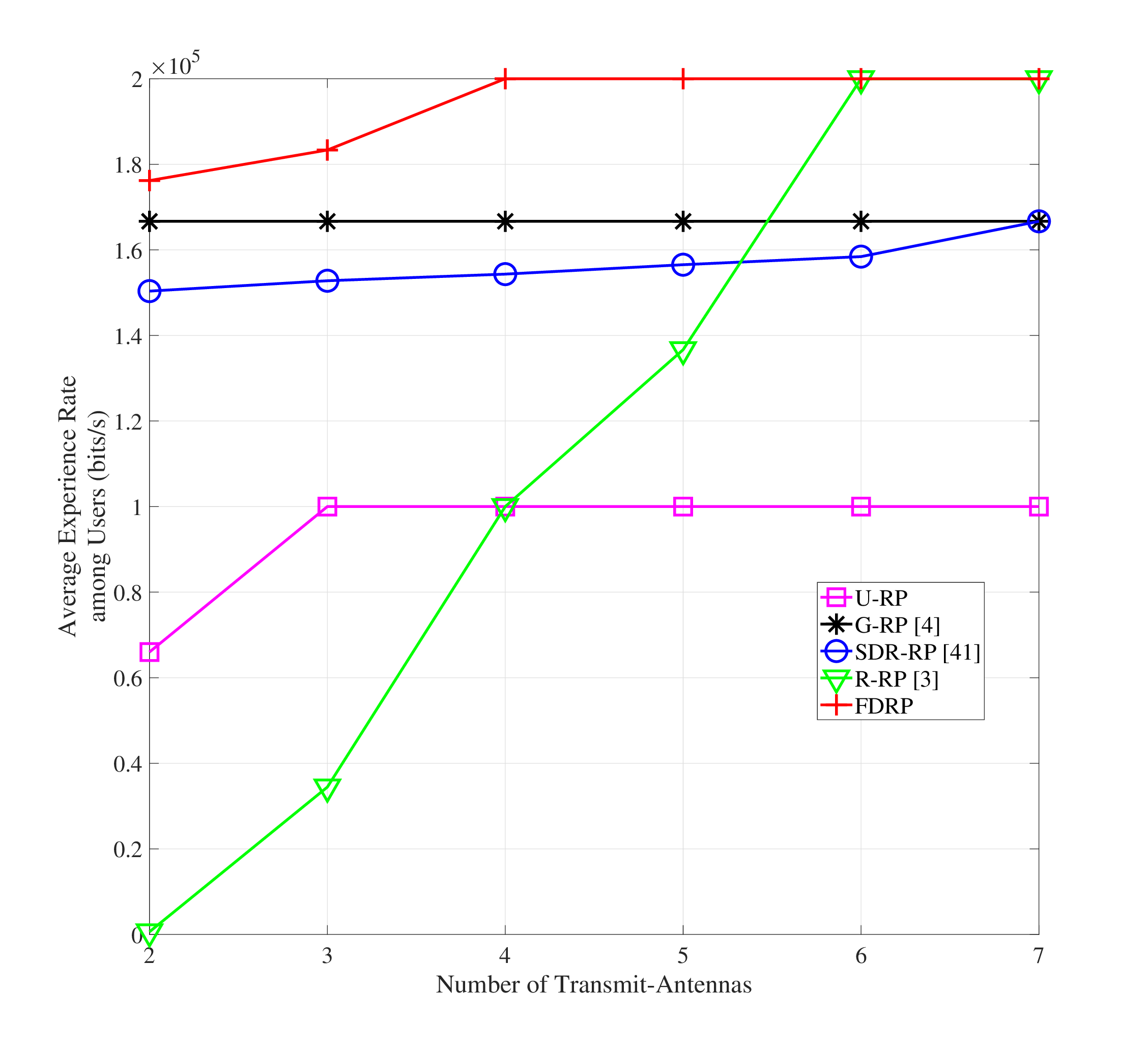}
    \caption{The impact of number of transmit-antennas on average experience rate among users. Transmission power budget is set to $P=-40$ dBm. Payload is set to $Q_k= 100$ bits, $\forall k \in \mathbb{K}$. 
    }
    \label{fig:differentantennas} 
\end{figure}

In Fig. \ref{fig:differentantennas}, the impact of number of transmit-antennas on the experience rate is demonstrated.
First, it can be seen that proposed FDRP algorithm obtains superior experience rate performance over others, under different numbers of transmit-antennas. In particular, with 4 or more antennas, all the users' payloads can be delivered in the first time slot, thus achieving the highest value of experience rate $2\times 10^5$ bits/s. 
Second, with a small number of transmit-antennas,  the subcarrier allocation indicator ($\eta$ in (6), cf. \cite{Bandi2020A}) of the R-RP algorithm tends to be small-valued, yielding poor sum-rate as well as experience rate performance. 
When the number of transmit-antennas is not smaller than the number of users, the constraint imposed on the subcarrier indicator by the R-RP algorithm can be ignored, yields an enhanced experience rate performance for the R-RP algorithm.
For the U-RP algorithm, it needs at least 3 time slots for a complete delivery when $N_t=2$. When $N_t$ keeps increasing, it begins to have feasible solutions for delivering its payload within 2 time slots.  Hence, its experience rate performance remains unchanged when $N_t \geq 2$ for the considered scenario.

\begin{figure}
	\centering
	\includegraphics[width=4.5 in]{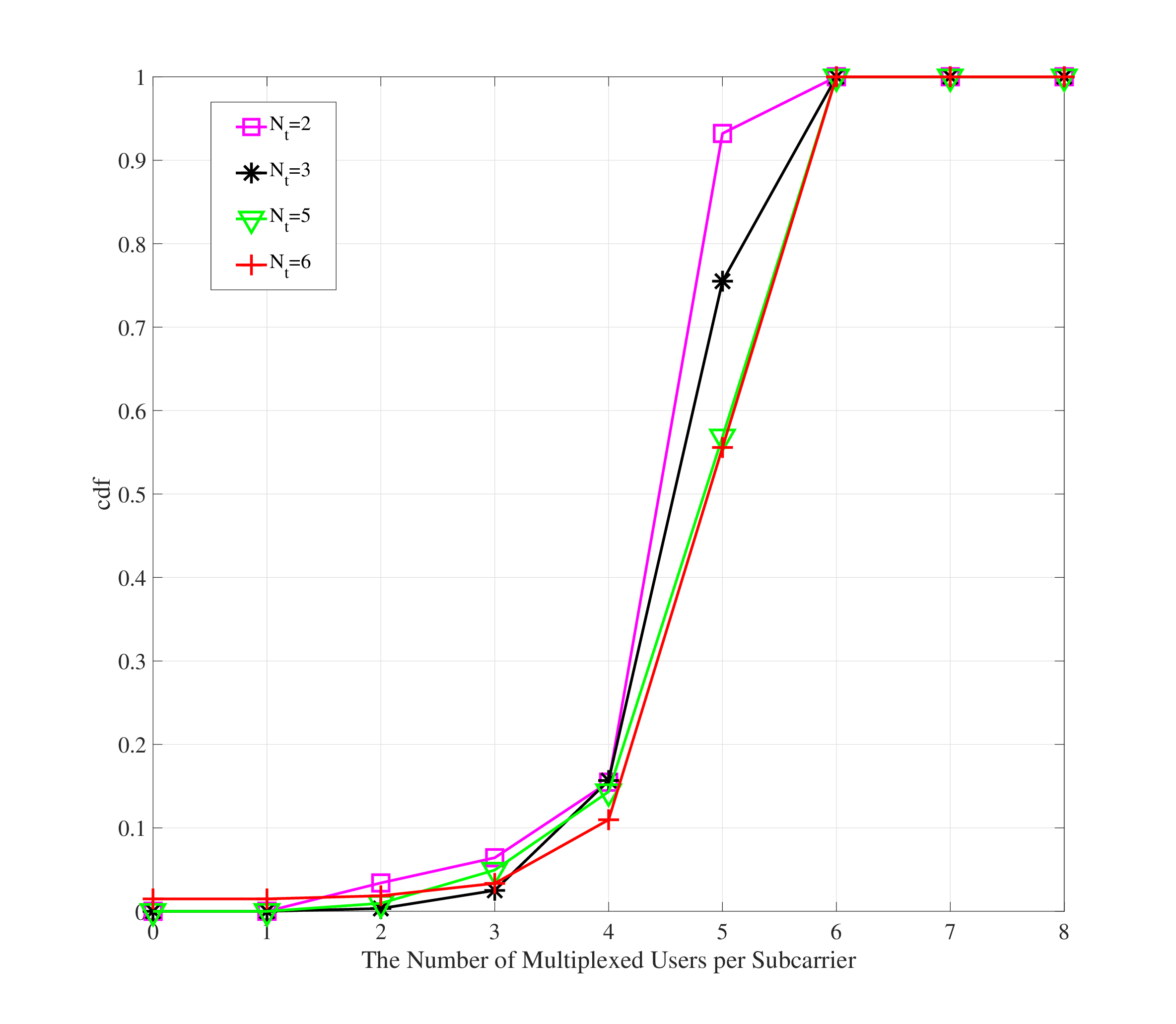}
    \caption{The behavior on the number of multiplexed users per subcarrier by the proposed algorithm. Transmission power budget is set to $P=-40$ dBm. Payload is set to $Q_k= 100$ bits, $\forall k \in \mathbb{K}$. 
    }
    \label{fig:Multiplexing} 
\end{figure}

In Fig. \ref{fig:Multiplexing}, the cumulative distribution function (cdf) of the number of the multiplexed users per subcarrier is demonstrated.
First, it shows that the proposed design can multiplex more users per subcarrier than the number of transmit-antennas.
Thus, the proposed design endorses higher DoF for  multiplexing users. As comparisons, the R-RP in \cite{Bandi2020A} strictly requires that the number of users multiplexed in no larger than the number of antennas, while the SDR-RP in \cite{Ku2015Joint} also tends to multiplex less users than the number of antennas for obtaining rank-1 solutions. 
Second, Fig. \ref{fig:Multiplexing} validates the conclusion of  Remark 4.  In practice, the number of users multiplexed on each subcarrier can not be arbitrarily large. For example, when $N_t=2$, it achieves up to 95\% percentage that, the number of the multiplexed users is not larger than 5.
Third, the algorithm tends to multiplex more users per subcarrier, with the increased number of transmit-antennas. It is obtained due to the increased DoFs at the transmitter side.

\section{Conclusion}
In this work, we have investigated a novel multistage dynamic programming problem for jointly optimizing time, frequency, space, and power resources.
Exploiting the unique property of structural sparsity in the resource allocated to multiple users, the multistage MINLP problem is first judiciously reformulated into a standard conic optimization.
Taking the max-min experience rate as the utility function,
a low-complexity full domain resource programming algorithm has been proposed. 
Aided by a dedicatedly designed SOCP problem for obtaining a feasible initial result, the proposed algorithm iteratively updates the allocation of full domains resource via a conic optimization, and converges to a near-optimum with fast convergence rate. 
Simulation results verify that, the proposed algorithm obtains significant performance enhancement over the benchmarks, while maintaining a reasonable level of computational complexity. Also, some interesting properties on the considered multistage MINLP problem has been discussed. 
The presented work in this paper offers a new viewpoint for multistage full domain resource allocation design, which holds the promise of 
exciting real-time media service in the years to come.

\end{document}